\documentclass[epj]{svjour}
\usepackage{epsfig} 
\usepackage{cite} 

\newcommand{\be}{\begin{equation}}
\newcommand{\ee}{\end{equation}}
\newcommand{\dd}{\mbox{\rm d}}
\newcommand{\mup}{\mbox{$\mu^-p$}}
\newcommand{\mud}{\mbox{$\mu^-d$}}
\newcommand{\pip}{\mbox{$\pi^-p$}}
\newcommand{\pbp}{\mbox{$\bar{p}p$}}

\newcommand{\eV}{\mbox{\rm eV}}
\newcommand{\xp}{\mbox{$x^-p$}}
\newcommand{\muxp}{\mbox{$\mu_{xp}$}}
\newcommand{\mumup}{\mbox{$\mu_{\mu p}$}}
\newcommand{\Ha}{\mbox{${\mathrm{H}}$}}   
\newcommand{\kstk}{\mbox{$k_\mathrm{Stk}$}}
   
\newcommand{\etal}{\mbox{\it et al.}}

\begin{document}  
\title{Collisional deexcitation of  exotic hydrogen atoms in highly excited states. 
II. Cascade calculations 
}
\titlerunning{Collisional deexcitation of exotic hydrogen atoms. II. Cascade calculations}
\author{ T.S. Jensen\inst{1,2} \and V.E. Markushin\inst{1}}
\institute{
Paul Scherrer Institute, CH-5232 Villigen PSI, Switzerland
\and 
Institut f{\"u}r Theoretische Physik der Universit{\"a}t Z{\"u}rich,
Winterthurerstrasse 190, CH-8057 Z{\"u}rich, Switzerland 
}
\date{Received: date / Revised version: date}
\abstract{
The atomic cascades in \mup\ and  \pbp\ atoms 
have been studied in detail using new results for the cross-sections of 
the scattering of highly excited exotic atoms from molecular hydrogen.  
The cascade calculations have been done with an updated version of 
the extended standard cascade model that computes the evolution in the
kinetic energy from the beginning of the cascade. 
The resulting X-ray yields, kinetic energy distributions, 
and cascade times are compared with the experimental data. 
}
\maketitle

\section{Introduction}
\label{SecIntro}
 
   The standard cascade model (SCM) of exotic hydrogen atoms, 
originally introduced by Leon and Bethe \cite{leon62} and later extended 
to include the evolution of the kinetic energy distribution 
during the atomic cascade \cite{markushin94,markushin99}, 
provides a fair description of many properties of 
atomic cascades, such as the X-ray yields  
and absorption fractions \cite{borie80,reifenrother89,aschenauer95,terada97}.   
However, no detailed calculations of the initial stage of the cascade 
were done until recently due to the lack of the corresponding  
cross-sections.  The upper stage of the cascade was commonly described 
with a phenomenological deexcitation mechanism, the so-called chemical 
deexcitation \cite{leon62}.  This mechanism was introduced   
in order to explain the data on cascade times, as the external Auger 
effect was found to be too slow at the initial stage of the cascade.  
The exact nature of the collisional deexcitation of highly excited exotic 
atoms remained a mystery for a long time until experimental studies of 
the initial stages of the atomic cascade became possible.  Recently 
 the energy distributions of \pip, \mup, and \mud\ 
atoms were studied with various time-of-flight methods 
\cite{crawford91,abbot97,badertscher01,kottmann99,pohl01}.  In particular, the measurements 
performed with \mup\ and \pbp\ 
at low density allow one to probe collisional deexcitation mechanisms 
for highly excited states as long as the energy distribution remains frozen 
during the lower cascade stage dominated by the radiative transitions.     

   In our recent paper \cite{jensen02last}, we studied the dynamics of collisional 
deexcitation of highly excited exotic atoms using the classical-trajectory 
Monte Carlo method.  The Coulomb transitions with large change of principal 
quantum number $n$ were found to be the dominant collisional deexcitation mechanism 
at high $n$, with the molecular structure of the hydrogen target 
being essential for the dominance of transitions with large $\Delta n$.        
The main goal of this paper is to investigate the atomic cascades in hydrogen-like 
atoms at low density and to confront the theoretical results  
with the experimental data on the X-ray yields, cascade times, and kinetic energy 
distributions for muonic and antiprotonic hydrogen.  
   As the lower part of the atomic cascade at low target densities is mainly 
dominated by the radiative deexcitation, the results of our calculations 
are only weakly affected by theoretical uncertainties in the collisional 
cross-sections for the low $n$ states. 
The detailed cascade calculations for high density targets will be published 
elsewhere.  

    The paper is organized as follows.  
The cascade model is described in Section~\ref{sect:ESCM}.  
The results of the cascade calculations for the $\mu^-p$ and $\bar{p}p$ atoms 
are presented in Section~\ref{sect:res}.
The conclusions are summarized in Section~\ref{sect:conc}.


\section{The extended standard cascade model}
\label{sect:ESCM}

\subsection{Overview}
\label{sect:overview}

   The extended standard cascade model (ESCM) is a kinetics model that 
includes all cascade processes of the standard cascade model 
\cite{leon62,borie80,terada97}: 
the radiative, Auger, and Coulomb deexcitation, Stark mixing, and, 
in case of hadronic atoms, nuclear absorption 
(see Table~\ref{table:processes} and references therein).      
In addition, the ESCM also takes into account the interaction between 
internal and external degrees of freedom of the exotic atom: 
the kinetic energy distribution changes during the cascade 
due to  the acceleration and deceleration mechanisms 
\cite{markushin94,markushin99,markushin02hyp,jensen02pin}.  
The new results for the collisional processes 
presented in~\cite{jensen02epjd,jensen02last}
are used in the  present version of the ESCM,  a significant  improvement over the 
previous calculations. 
\begin{table*}
\begin{center}
\begin{tabular}{|ll|c|}\hline
   Process &  Example  &  Reference  
\\ \hline\hline
Stark mixing: & $(x^-p)_{nl_i}+\mathrm{H}_2\to (x^-p)_{nl_f}+\mathrm{H}_2^*$  &  
                \cite{jensen01,jensen02hyp,jensen02epjd,jensen02last} 
\\ \hline
External Auger effect: & $(x^-p)_i+\mathrm{H}\to (x^-p)_f+p+e^-$  & 
                \cite{bukhvostov82,jensen02last} 
\\ \hline
Coulomb deexcitation: & $(x^-p)_{n_i}+\mathrm{H}_2\to (x^-p)_{n_f}+\mathrm{H}_2^*,\; n_f<n_i$  &  
                \cite{jensen02hyp,jensen02last}
\\ \hline
Elastic: & $(x^-p)_{nl}+\mathrm{H}_2\to (x^-p)_{nl}+\mathrm{H}_2^*$  & 
                \cite{jensen02epjd,jensen02last} 
\\ \hline
Absorption: & $(\pi^-p)_i+\mathrm{H}\to \pi^0 +n +\mathrm{H}$ & 
                \cite{jensen01,jensen02hyp,jensen02epjd}  
\\ \hline
Radiative: & $(x^-p)_{n_il_i}\to (x^-p)_{n_fl_f}+\gamma$  &  
                \cite{bethe57} 
\\ \hline
Nuclear reaction: &  $(\pi^-p)_{ns}\to \pi^0 +n,\;\gamma+ n$ &  
                \cite{leon62} 
\\ \hline
Weak decay: &  $\pi^-\to \mu^-\bar{\nu}_\mu\nu_e$  &  
\\  \hline
\end{tabular}
\end{center}

\caption{\label{table:processes}
Processes included in the extended standard  cascade model.}
\end{table*}

The cascade in exotic atoms is divided into two parts: 
the classical domain for high $n$ 
and the quantum mechanical domain for low  $n$. The
\xp\ can be described classically for quantum numbers $n\gg1$, and for convenience 
we use the properties of the 
Auger deexcitation to define the $n$-ranges of the  two domains. The Auger deexcitation
rates are known to have a maximum for $n=n_c$ where $n_c$ is the largest $n$ for
which  the binding energy released in a $\Delta n=1$ transition is enough to ionize
the $\Ha_2$ molecule~\cite{leon62,bukhvostov82,jensen02last}: $n_c=7$ for \mup\ and
$n_c=12$ for \pbp. Our calculations
show that the Coulomb deexcitation dominates the Auger deexcitation
for $n>n_c$ \cite{jensen02last}. We
 define the classical domain by the conditions  $n>n_c$.
The classical domain, where the  processes were calculated in the classical Monte Carlo (CMC)
 model  with the molecular structure  of the target taken into account, will be
discussed in Section~\ref{sect:includecmc}. In the quantum mechanical domain, $n\leq n_c$, 
the close-coupling method and  semiclassical approximations~\cite{jensen99,jensen01,jensen02epjd}
 were used
to calculate the  differential cross-sections, $\dd\sigma/\dd \cos\theta$,
 of the collisional processes 
\be
 (x^-p)_{nl} + \Ha \to  (x^-p)_{nl'} + \Ha   
\ee
and the cross-sections for absorption during collision and Auger transitions
\begin{eqnarray}
      (x^-p)_{nl} + \Ha    & \to &  
      \mathrm{absorption}   ,
\\
      (x^-p)_{n_il_i} + \Ha  & \to &  
      (x^-p)_{n_f} + p+e^- 
\end{eqnarray} 
in an energy range relevant for atomic cascade.
The new calculations allow us to describe the competition between deceleration and Stark 
mixing, as well as the absorption in hadronic atoms due to Stark collisions,   
without employing any fitting parameters related to the Stark mixing and 
deceleration, like \kstk\ used in many earlier calculations~\cite{borie80}. Our
 cascade code 
makes full use of the  differential cross-sections so that 
the kinetics is treated more
accurately than in cascade models using continuous deceleration~\cite{koike02hyp}.

The differential cross
sections used in the present cascade model in the case of muonic hydrogen
were calculated in the fully quantum mechanical 
close-coupling  framework for $n=2-5$ and in the semiclassical approximation
for $n=6-7$.  The statistically weighted differential cross-sections were used
for $n\geq 6$. In antiprotonic hydrogen, the fully quantum mechanical
results are not yet available below the $ns$ thresholds. Therefore, we used 
the results of the semiclassical model in the range $n=2-9$ and the
fixed field model for $n=10-12$.

Another approximation used in this paper
 is related to   the Auger deexcitation because the
eikonal approximation~\cite{jensen02last}
 does not give the differential cross-section and the distribution
over final $l$. Here we use the  differential cross-section for the $l$-average
Stark and elastic transitions and a statistical distribution over final $l$.  


   Concerning the acceleration mechanisms, the present calculations include  
Coulomb deexcitation through the whole cascade.  For high $n$ states, 
the classical-trajectory Monte Carlo method~\cite{jensen02last} was used
to obtain the cross-sections of the inelastic collisions   
\begin{eqnarray}
      (x^-p)_{n_i} + \mathrm{H}_2  & \to &  
      (x^-p)_{n_f} + X  ,\;
      X = \mathrm{H}_2,\; \mathrm{H}_2^*,\; \mathrm{H}+\mathrm{H}  . 
\end{eqnarray} 
For low  $n$, where  the 
classical-trajectory method cannot be expected to give  reliable results, 
we parameterize the Coulomb deexcitation  cross-section  as follows\footnote{The
$n=2$ state in muonic hydrogen is, however, treated differently. Coulomb deexcitation  
is assumed only to take place in the $2s$ state and only when the kinetic  energy is below 
the $2p$ threshold. In this case we use the measured 
 Coulomb deexcitation rate
$\lambda_{2s\to 1s}^\mathrm{Coulomb}=4.4\cdot 10^{11}N\;\mathrm{s}^{-1}$~\cite{pohl01} where
$N$ is the density of the target in units of liquid hydrogen density
(LHD).}
\be
 \sigma_{n\to n-1}(E_\mathrm{cm})=c\frac{n^{\gamma}}{E_\mathrm{cm}}\cdot 
  \frac{\mu_{\mu p}}{\mu_{xp}}
  \label{coulfit}
\ee
where $E_\mathrm{cm}$ is the CMS kinetic energy and \mumup\ and \muxp\ are the reduced masses
of the \mup\ and the \xp, respectively. We use the values 
\begin{eqnarray}
  c&=&1.2\cdot 10^{-3} a_0^2 \;\eV ,\\
  \gamma&=& 3.5 
\end{eqnarray}
which  gives a fair description of the  Coulomb cross-sections of Bracci 
and Fiorentini~\cite{bracci78} for $n\le 7$.

Other calculations~\cite{ponomarev99mup,ponomarev99,kravtsov99} predict 
significantly smaller Coulomb deexcitation rates than \cite{bracci78}.
However, the rapid increase
in  the Coulomb  cross-sections for energies approaching zero is predicted by all models.
The dependence on the reduced mass is based on  an estimate in~\cite{bracci78} and
our classical Monte Carlo calculations confirm that it is a fair approximation. 
We assume, furthermore, that only $\Delta n=1$ transitions are important at low $n$, 
the distribution over final $l$ is statistical, and the angular distribution is isotropic. 
As long as the mechanism responsible for Coulomb deexcitation at low $n $ is not fully
understood\footnote{The investigation of the  role of 
molecular resonances~\cite{wallenius96,jonsell99} is beyond the scope of this study.},  
this process enters as 
a major uncertainty in calculations of the kinetic energy evolution at low $n$.
In this paper we restrict the cascade calculations to observables that are not very sensitive 
to the Coulomb deexcitation at low $n$. In the case of kinetic energy distributions
this usually means low densities. Other observables, like the X-ray yields in \mup\ and
cascade times, are less sensitive to the Coulomb deexcitation at low $n$.

\begin{figure}
\center{
\epsfig{file=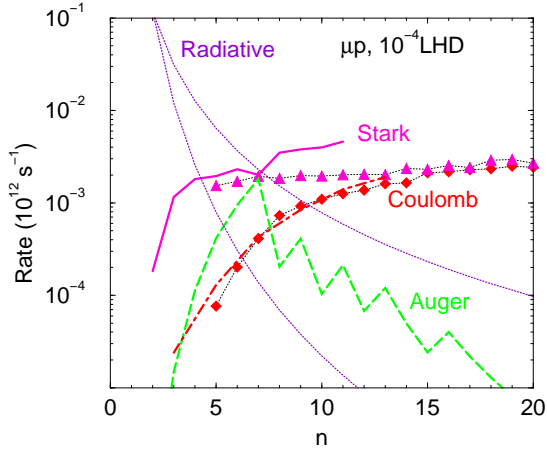, width=7.1cm}
}
\caption{\label{fig:muprates}
The  $l$-average rates at $T=1\;$eV for
 muonic hydrogen  
 in gaseous target at $10^{-4}\;$LHD.
The Coulomb deexcitation (filled diamonds) and Stark mixing 
(filled triangles)
 rates calculated
in the classical-trajectory model~\protect\cite{jensen02last} 
are shown in comparison with the results of
 the semiclassical fixed field model for Stark mixing (solid line), 
 Auger deexcitation (light dashed line),
and nuclear absorption during collisions (dashed line)~\cite{jensen02epjd,jensen02last}. 
The Coulomb deexcitation rate from~\protect\cite{bracci78}
is shown with a dash-dotted line.
 The radiative $np\to 1s$ and $n(n-1)\to (n-1)(n-2)$ rates are shown with dotted lines.
}
\end{figure}

\begin{figure}
\center{
\epsfig{file=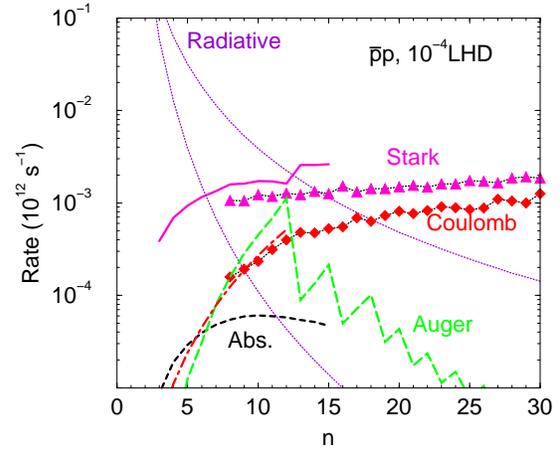, width=7.1cm}
}
\caption{\label{fig:pbprates}
The  $l$-average rates at $T=1\;$eV for
 antiprotonic hydrogen  
 in gaseous target at $10^{-4}\;$LHD.
The Coulomb deexcitation (filled diamonds) and Stark mixing 
(filled triangles)
 rates calculated
in the classical-trajectory model~\protect\cite{jensen02last} 
are shown in comparison with the results of
 the semiclassical fixed field model for Stark mixing (solid line), 
 Auger deexcitation (light dashed line),
and nuclear absorption during collisions (dashed line)~\protect\cite{jensen02epjd,jensen02last}. 
The Coulomb deexcitation rate from  Equation~(\ref{coulfit}) is shown with a dash-dotted line.
 The radiative $np\to 1s$ and $n(n-1)\to (n-1)(n-2)$ rates are shown with dotted lines.
}
\end{figure}

Figures~\ref{fig:muprates} and~\ref{fig:pbprates}  show an overview of the cascade rates
at $10^{-4}\;$LHD  in muonic and
antiprotonic hydrogen calculated in the classical-trajectory model with
molecular target  and in 
semiclassical  approximations with atomic target. The absorption rate in
antiprotonic hydrogen shows only the absorption from the $ns$ states
during collisions. Cascade calculations show, in agreement with the 
experimental results~\cite{bendiscioli01}, that absorption at densities below
$10^{-3}\;$LHD takes place mainly from the $p$ states.  

\subsection{Classical-trajectory Monte Carlo model in the cascade calculations}
\label{sect:includecmc}

The  classical-trajectory  Monte Carlo calculations have been  included in the
cascade calculations by using two different methods. 
The earlier versions of the cascade model~\cite{markushin02hyp}
 was based on the explicit
use of the cross-sections~\cite{jensen02last} with the following initial
 conditions.   
A discrete set of 9 initial kinetic energies    
in the interval $0.05\;\mbox{eV}\leq T \leq 20\;\mbox{eV}$ was used, 
and the initial principal quantum numbers were taken in the range 
$8 \leq n_i \leq 20$ for \mup\ and  $13 \leq n_i \leq 35$ for \pbp.   
For each combination of laboratory kinetic energies $T$ and 
principal quantum numbers $n_i$, the classical  trajectories were calculated 
using the statistical distribution in quantum number $l_i$ as the initial 
condition.
The differential cross-sections $\mathrm{d}\sigma/\mathrm{d}\cos\theta$ were
calculated separately for the three groups of reactions:
\be
  n_i l_i\to \left\{
\begin{array}{ll}
  n_il_i&\mathrm{elastic}\\
  n_il_f\; (l_f\neq l_i)&\mathrm{Stark}\\
  n_fl_f\; (n_f< n_i)&\mathrm{Coulomb}
\end{array}\right.
\ee
The effect of excitation and dissociation of the hydrogen molecule in
 Coulomb deexcitations were taken into account by distributing 
 the final kinetic energies of the three
atoms ($x^-p$ and two H) in the CMS  according
to the classical phase space. This approximation agrees fairly well 
with the calculations \cite{jensen02last}.

More detailed initial and final states are used in the present cascade model
 where several trajectories (21 for \mup\ and 6 for \pbp)
 were calculated for each combination of the quantum numbers
$n_i$ (in the range given above) and $l_i=0,..,n_i-1$, and  
31 kinetic energy points in the range $0.3-32\;$eV. 
In muonic hydrogen this gives a total of 118482 trajectories.
The range of impact parameters was divided into three intervals, 
$(0;2a_0)$, $(2a_0;4a_0)$, and $(4a_0;7a_0)$, with a third of the 
collisions in each and the trajectories  uniformly distributed in $\rho^2$.  
For each trajectory, the final quantum numbers
$n_f$ and $l_f$, and the final laboratory kinetic energy $T_f$ were stored.  
The resulting lists of transitions in  the form
\be
  n_i\; l_i\; T_i \;\to\; n_f\; l_f\; T_f,
\ee 
with the weights of the impact parameter intervals properly taken into account, 
were used directly by the cascade program by redirecting  the final state to the
initial state of the next collision.
We consider this method  as more accurate than the earlier used method 
for cascade calculations in the upper part of the cascade.

\subsection{Initial conditions}
\label{sect:init}

The initial conditions for the cascade calculations are defined by
the initial distributions in
the quantum numbers $n$ and $l$ and the laboratory kinetic energy $T$ of the \xp. 
In the simplest picture, the $x^-$ is captured by the proton after ionization
of the hydrogen atom in a state with large overlap with the $1s$ electron
wave function; this gives $n_i\sim\sqrt{\muxp/m_e}$ ($\sim 14$ for
muonic and $\sim 30$ for antiprotonic hydrogen).
More elaborate calculations, which takes the molecular structure of the target
into account, predict distributions in the initial $n$ which peak
at lower values: for example, the maximum at $n=11$ for muonic hydrogen \cite{cohen99}. 
A shift towards lower initial $n$ values is also found in \cite{korenman96}.
The new experimental data on muonic
hydrogen kinetic energy distributions from PSI are sensitive to the initial
distributions as discussed below in Section~\ref{sect:res:mup:kin}.

Unless otherwise indicated, we use the  initial  principal quantum number
$n_i=14$ for \mup\ 
and $n_i=30$  for \pbp\
 and the initial kinetic  energy given by a Maxwell distribution 
\be
  w(T_i)=3\sqrt{\frac{3T_i}{2\pi T_0}}\exp\left(-\frac{3T_i}{2\pi T_0}\right)
  \label{tinit}
\ee 
with $T_0=0.5$~eV. In all cases the 
$l_i$ distribution is  statistical.
  

\section{Results}
\label{sect:res}

\subsection{Muonic hydrogen}
\label{sect:mup}
The cascade calculations in muonic hydrogen have been done in the density range
$(10^{-8}-1)\;$LHD using various initial conditions. The results are compared with
the experimental data for the metastable $2s$ population, X-ray yields, 
cascade times, and kinetic energy 
distributions~\cite{anderhub77,egan81,anderhub84,sakamoto99,bregant98,lauss98,pohl01,pohl02hyp,kottmann02hyp}.

\subsubsection{The metastable $2s$ fraction}

The metastable $2s$ state of muonic hydrogen  is of great interest 
because  it allows, under certain conditions, to measure  the \mup\ Lamb shift, $2s-2p$,
from which the proton charge radius can be determined with
high precision~\cite{taqqu99}. The feasibility of this experiment, presently
in progress at PSI~\cite{kottmann02hyp},
relies on a sizeable fraction of muonic hydrogen atoms in the $2s$ state  
with a sufficiently long lifetime. In the absence of collisions, the lifetime of 
the $2s$ state is determined by the muon lifetime ($\tau_{\mu}=2\cdot 10^{-6}\;$s) as  
the  rate for  the $2s\to 1s$ two-photon transition is negligibly small. 
At target densities above $10^{-7}\;$LHD, the lifetime of the $2s$ state depends on the
kinetic energy of the \mup. Below the $2p$ threshold, $T_{2p}=0.3\;$eV, the $2s$ state
is long-lived because $2s\to 2p$ Stark transitions followed by the fast radiative
deexcitation  $2p\to 1s$ are energetically forbidden. The induced radiative transition can,
however, occur during collisions, see \cite{mueller75,cohen81} and below. 
For kinetic energies above the $2p$ threshold the collisions
with the target molecules lead to a  competition between 
deceleration  and depletion via $2s\to 2p$ Stark transitions followed 
by the $2p\to 1s$ radiative transition.
The {\it metastable $2s$ fraction}, $R_{2s}$,
 is defined
as the fraction of all created \mup\ atoms which end up in the $2s$ state with 
kinetic energies below $T_{2p}$. 

Before we present the results of our full cascade model, it is illuminating to 
consider the problem of the metastable $2s$ fraction in a simplified framework.
The  metastable $2s$ fraction consists of the
\mup\ that are formed with kinetic energies below $T_{2p}$
 and those that are formed with kinetic energies 
above $T_{2p}$ but slow down below threshold due to  elastic collisions.  
The median kinetic energy at 1~mbar is 1.5~eV~\cite{pohl01}, so the deceleration
process is important.
This leads to the definition of  the {\it surviving metastable  fraction}, $f(T)$, 
which is the probability
that the \mup\ atom in the $2s$ state with initial kinetic energy $T$ reaches 
energies below the $2s$ threshold by slowing down in elastic collisions~\cite{jensen99}. 

In the approximation of continuous energy loss the evolution in the kinetic energy 
is given by
\be
  -T^{-1}\frac{\dd T}{\dd t}=2\frac{M_{\mu p} M_\mathrm{H}}{(M_{\mu p} + M_\mathrm{H})^2}
  N v \sigma_{2s\to 2s}^\mathrm{tr}(T)
 \label{eq:tevolution}
\ee
where $M_{\mu p}$ and $M_\mathrm{H}$ are the masses of the \mup\ and the H, respectively,
 $v$ the velocity of the \mup, and $\sigma_{2s\to 2s}^\mathrm{tr}(T)$ the
transport cross-section~\cite{jensen02epjd}.
Equation~(\ref{eq:tevolution})
was used by Carboni and Fiorentini~\cite{carboni77} to get
the following estimate of the surviving metastable 
fraction
\begin{equation}  
f(T)=\exp\left(-\frac{(M_{\mu p} +M_\mathrm{H})^2}{2M_{\mu p}M_\mathrm{H}}
 \int_{T_{2p}}^{T}\frac{\sigma_{2s\to 2p}(T^\prime)}{T^\prime 
  \sigma_{2s\to 2s}^\mathrm{tr}(T^\prime)}\dd T^\prime\right) .
\label{formula}
\end{equation}
It was found that a sizeable fraction of $(\mup)_{2s}$ atoms formed at 
kinetic energies below 1.3 eV slows down below threshold.

To provide a more realistic treatment
of  the evolution in kinetic energy we use a Monte Carlo program 
based on the differential cross-sections for the four processes
$2s\rightarrow 2s$, $2s\rightarrow 2p$, $2p\rightarrow 2s$ and  $2p\rightarrow 2p$. 
In addition to the collisional 
processes, the  $2p\rightarrow 1s$ radiative transition is also included in the 
calculation. The fate of a \mup\ formed in the $2s$ state with kinetic energy $T$ 
is thus either to undergo $2p\rightarrow 1s$ radiative transition after the Stark mixing 
$2s\rightarrow 2p$ or to end up in the $2s$ state with kinetic energy 
below the threshold with probability $f(T)$.  

\begin{figure}
\center{
\epsfig{file=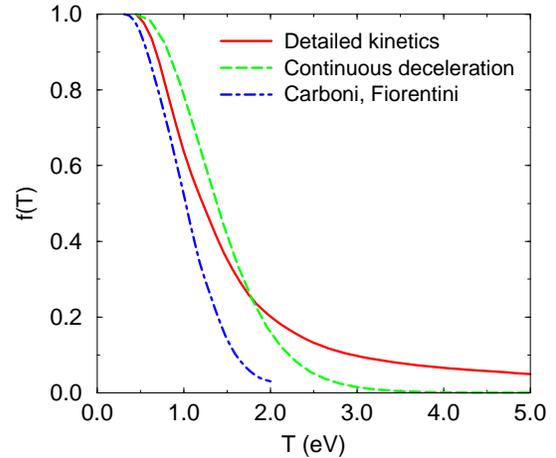, width=7.1cm}
}
\caption{\label{fig:f} 
 Energy dependence of the surviving metastable  
fraction $f(T)$ of the $2s$ state of muonic hydrogen. The result of the detailed kinetics
 calculations (solid line) is shown in comparison with the 
 approximation of continuous deceleration (\ref{formula}) (dashed line) 
 and the result of Carboni and Fiorentini~\protect\cite{carboni77}.
}
\end{figure}

Figure~\ref{fig:f} shows the surviving metastable 
fraction, $f(T)$, calculated with 
the Monte Carlo program for target density $10^{-6}\;\mathrm{LHD} < N < 10^{-2}\;\mathrm{LHD}$.  
The approximation~(\ref{formula}) gives somewhat higher values for  
the survival probability than the exact kinetics calculation at $T<1.4\;$eV.  
The Monte Carlo results at high energies ($T>1.5\;$eV) are significantly  
larger than those obtained from Equation (\ref{formula}) where continuous 
energy loss is assumed. 
The reason is that  there is a significant contribution
to $f(T)$ from large-angle scattering.
The result of~\cite{carboni77} is considerably smaller than the present result
due to differences in the cross-sections and the fact that in \cite{carboni77}
the molecular hydrogen mass was used instead of the atomic one in order to get the
correct value for the threshold. By using  the surviving metastable
fraction $f(T)$, the measured kinetic energy distributions on arrival in the
$1s$ state~\cite{pohl01}, and the  $2s$ arrival probability\footnote{The $2s$ arrival
probability is the
fraction of formed exotic atoms that arrives  at the $n=2$ level in the $2s$ state.
 It can be estimated from measurements of $K$ X-ray yields 
 using the known $Y_{np\to2s}/Y_{np\to 1s}$ branching ratios.}~\cite{anderhub84}, 
one finds the metastable $2s$ fraction to be $\sim 1$\% in the 
pressure region $0.06-16$~mbar as discussed in \cite{pohl01}.

After reaching thermal energies the $(\mup)_{2s}$ atoms may undergo radiative deexcitation
during collisions because of the Stark effect~\cite{mueller75,cohen81}:
\be
 (\mup)_{2s}+\Ha\to (\mup)_{2s-2p}+\Ha\to (\mup)_{1s}+\gamma(1.9\;\mathrm{keV})+\Ha .
 \label{2squenching:rad}
\ee
The radiative quenching of  thermal $(\mup)_{2s}$ atoms would  result in delayed
$K_\alpha$ X-rays. However, the experiments~\cite{anderhub77,egan81,anderhub84} have 
not observed this effect. This apparent contradiction  was recently resolved by
the \mup\ diffusion  experiment~\cite{pohl01, pohl02hyp} where 0.9~keV
$(\mup)_{1s}$ atoms, resulting from the Coulomb deexcitation process
\be
 (\mup)_{2s}+\Ha\to (\mup)_{1s}(0.9\;\mathrm{keV})+\Ha(1.0\;\mathrm{keV}),
 \label{2squenching:nonrad}
\ee
were found. The measured   non-radiative quenching rate is about $2\;\mu\mathrm{s}^{-1}$
 at 1 mbar~\cite{pohl01, pohl02hyp} which
is more than 
an order of magnitude larger than the theoretical predictions for the  radiative quenching
rate~\cite{mueller75,cohen81}.

\begin{figure}
\center{
\epsfig{file=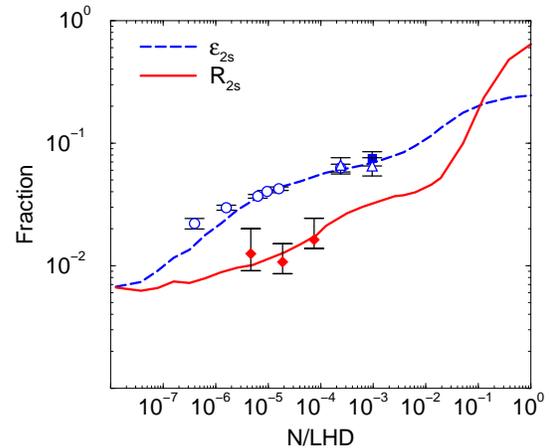, width=7.1cm}
}
\caption{\label{fig:mup2s}  
{\it Ab initio} cascade calculation of the $2s$ arrival probability, $\epsilon_{2s}$ 
(dashed line),
in comparison with the values derived  from X-ray 
measurements:~\protect\cite{anderhub84} (circles),
\protect\cite{anderhub77} (triangles), and \protect\cite{egan81} (filled square). The solid line shows the
the metastable $2s$ fraction, $R_{2s}$; the corresponding experimental data (filled diamonds)
 are from~\protect\cite{pohl01}.
}
\end{figure}

The $2s$ arrival probability, $\epsilon_{2s}$, and 
metastable $2s$ fraction calculated in the ESCM
with the initial conditions specified in Section~\ref{sect:init}
is shown in Figure~\ref{fig:mup2s}. 
The calculated $2s$ arrival probability is in good agreement with the experimental
data obtained from measurements of X-ray yields~\cite{anderhub77,anderhub84,egan81}.
The arrival probability $\epsilon_{2s}$ increases steadily with increasing 
density from 1\% at $10^{-7}\;$LHD to
 25\% in liquid hydrogen. This behavior has a simple qualitative explanation: at the 
lowest  densities the cascade proceeds purely radiatively from high $n$ values. 
As only $\Delta l=1$ transitions are allowed, the circular states are predominantly 
populated for low $n$. The  majority
of exotic atoms, therefore, go through the radiative $2p\to 1s$ transition without reaching
the $2s$ state.  With the density increasing the Stark mixing becomes efficient 
at lower $n$ levels and more $np\to 2s$ radiative transitions take place. In liquid hydrogen,
$\epsilon_{2s}$ is given by the statistical weight of the $2s$ state.

In the density range $10^{-7}-10^{-4}\;$LHD, the calculated metastable fraction
increases from 0.7\% to 2\%, and this  is 
 in  perfect agreement  with the  experimental result~\cite{pohl01} 
obtained in  measurements of 0.9~keV $(\mup)_{1s}$ atoms.
The metastable fraction grows slowly with density until about $0.01\;$LHD. Above $0.01\;$LHD
the  growth becomes faster and  $R_{2s}$  reaches
65\% in liquid hydrogen. 
The large metastable fraction at high densities is a consequence 
of the dominance of  the   slowing down in $2l_i\to 2l_f$ collisions over
the radiative $2p\to1s$ deexcitation.


\begin{figure}
\center{
\epsfig{file=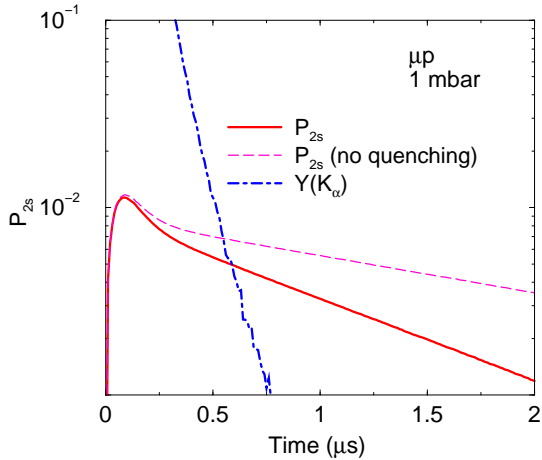, width=7.1cm}
}
\caption{\label{fig:mup2stime} 
The time dependence of the $2s$ population, $P_{2s}$ (solid line), of muonic hydrogen
at 1~mbar. The thin dashed line shows $P_{2s}$ calculated
without collisional quenching of the metastable $2s$ state. The dashed-dotted line
shows the distribution of  the $K_\alpha$ X-rays in units of $\mu\mathrm{s}^{-1}$. 
}
\end{figure}

Figure~\ref{fig:mup2stime} shows the time dependence of the 
$2s$ population, $P_{2s}$, at 1~mbar. The $2s$ population grows rapidly with time 
in the beginning of the cascade and reaches a maximum of 1.2\% after $0.1\;\mu$s.
  The development at later times clearly shows the two components of $P_{2s}$.
The short-lived component  consists of the \mup\ atoms with  kinetic energies above
the threshold, 
$T_{2p}$, which  go through the $2s\to 2p$ Stark transition followed by the radiative deexcitation.
The long-lived component consists of the metastable $(\mup)_{2s}$. 
The atomic cascade is essentially terminated after $0.7\;\mu$s: the \mup\ atoms are either
in the $1s$ state or the metastable $2s$ state. The metastable $2s$ state is quenched
non-radiatively~(\ref{2squenching:nonrad}) with a rate at 1 mbar of 
about 4 times the rate for muon decay. 

The results in Figure~\ref{fig:mup2stime} were calculated under the assumption that all
\mup\ atoms were formed at the same time. In reality,
the muons are injected into the target, slow down in collisions with the $\Ha_2$ molecules,
and are then captured by a proton. 
Having to go through the  slowing down process  means that the time of formation is given by a density 
dependent distribution --- the so-called stop time distribution. 
The measured stop time at  1~mbar is $\tau_\mathrm{stop}=0.48\pm0.13\;\mu$s~\cite{pohl01}
 which, when taken into account,
does not change the overall features of Figure~\ref{fig:mup2stime} except for the time offset. 

The results for the $2s$ population are extremely important for the $2s-2p$ Lamb shift 
experiment~\cite{taqqu99,kottmann02hyp}. In this  experiment,  a laser will be used 
 to induce $2s\to 2p$ transitions which are followed by the fast radiative $2p\to 1s$
deexcitation. The $K_\alpha$ X-rays are detected in coincidence
with the laser pulse and the $2s-2p$ energy splitting is then deduced from
the  laser wavelength corresponding to the strongest $K_\alpha$ signal. 
This laser experiment is feasible provided  the $2s$ population is not too small  and 
the $K_\alpha$ background produced by the cascade is negligible. These conditions are fulfilled
at very low densities (around a few mbar) after a short delay time as discussed above.

\subsubsection{X-ray yields}

\begin{figure}
\center{
\epsfig{file=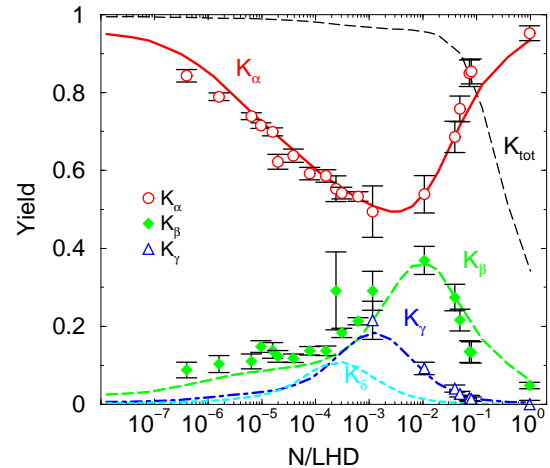, width=7.1cm}
}
\caption{\label{fig:mup:xray}
The density dependence of the relative X-ray yields,  
$K_{\alpha}$, $K_{\beta}$, and $K_{\gamma}$,  
and the absolute total yield $K_{\mathrm{tot}}$
in muonic hydrogen. 
The experimental data are from \protect\cite{anderhub84,bregant98,lauss98}.  
}
\end{figure}
\begin{figure}
\center{
\epsfig{file=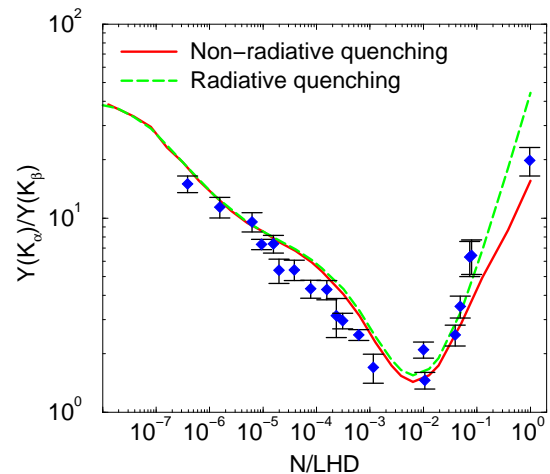, width=7.1cm}
}
\caption{\label{fig:mup:kakb}
The ratio of the $K_{\alpha}$ and $K_{\beta}$ yields 
for the non-radiative (solid line) and radiative (dashed line) quenching 
of the $2s$ state.    
The data are from \protect\cite{anderhub84,lauss98,bregant98,sakamoto99}. 
}
\end{figure}

The calculated relative X-ray yields for the \mup\ atom are shown in 
comparison with the experimental data in Figure~\ref{fig:mup:xray}. 
The agreement between theory and experiment
is good for the   $K_{\alpha}$, $K_{\beta}$, and $K_{\gamma}$ yields.
The  relative
$K_\alpha$ yield is higher than 90\% at $10^{-7}\;$LHD, it decreases with the density increasing
and reaches a minimum of 50\% at $2\cdot 10^{-3}\;$LHD. Above  $2\cdot 10^{-3}\;$LHD
the relative $K_\alpha$ yield increases and ends at 94\% in liquid hydrogen. 
The opposite behavior is seen in the other $K$ yields. They start out 
close to zero at $10^{-7}\;$LHD, increase with the density and reach their
maxima (at $0.01\;$LHD for $K_\beta$, $10^{-3}\;$LHD for  $K_\gamma$, and
$3\cdot 10^{-4}\;$LHD for $K_\delta$), and, with the exception of 
$K_\beta$, practically vanish at LHD.  The density dependence of the yields
can be explained qualitatively
in a way similar to the discussion of the $2s$ population above.
At $10^{-7}\;$LHD the cascade is almost purely radiative and proceeds at low
$n$ through the circular states, so that almost all \mup\ atoms go through the
radiative $2p\to 1s$ transition whereas the transitions 
$np\to 1s$ with $n>2$ are much weaker. With increasing density the Stark
mixing becomes more efficient and leads to higher populations of the 
$np$ states with $n>2$. This explains the increase in the relative
$K_{\geq\beta}$ yields and consequently also the decrease in $K_\alpha$. 
The decrease in the $K_{\geq\beta}$ yields for higher densities is due
to the collisional deexcitation $n\to n-1$  which begin
 to dominate the radiative transitions. 

At high density, a significant fraction of the \mup\ atoms slows down at 
the very end of the cascade before reaching the ground state and 
populate the metastable $2s$ state with high probability 
($R_{2s}\approx 0.65$ in liquid hydrogen) as discussed above.  
This makes the ratio of $K_{\alpha}$ and $K_{\beta}$ yields in liquid 
hydrogen very sensitive to a quenching mechanism of the $2s$ state.    
As demonstrated in Figure~\ref{fig:mup:kakb}, the X-ray data favor 
non-radiative quenching~(\ref{2squenching:nonrad}) in agreement with the recent experiment 
on direct search for the metastable $2s$ state \cite{pohl01}.   
It would be very important to check this result directly by measuring the 
{\it absolute} X-ray yield in liquid hydrogen: the non-radiative quenching leads to 
a significant suppression of the X-ray yield: $Y(K_{\mathrm{tot}}) \approx 0.35$. 

\subsubsection{Cascade time}

\begin{figure}
\center{
\epsfig{file=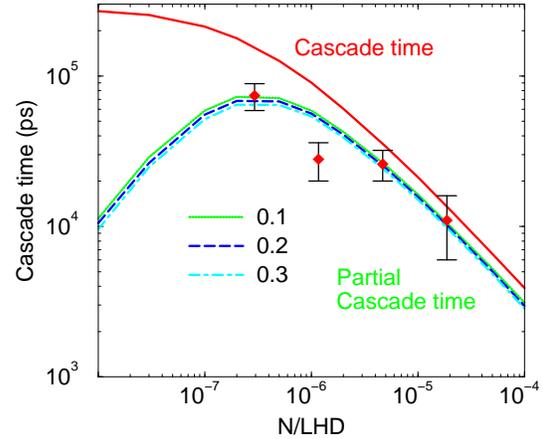, width=7.1cm}
}
\caption{\label{fig:mup:time} 
The calculated cascade time and the partial cascade time in \mup\ vs. density. 
The partial cascade
times are shown for different values of the parameter $\delta$. 
The experimental data are from \protect\cite{pohl01}.  
}
\end{figure}

The cascade time in muonic hydrogen at low pressures (0.25-16 mbar) was measured
recently in the $(\mup)_{1s}$ diffusion experiment at PSI \cite{pohl01}. 
The experimentally measured value is the so-called 
{\it partial cascade time} that  is defined as
the time between formation of the \mup\ and the 
last ``significant'' collision during the cascade,
and   is, therefore,  smaller than  the cascade time  usually defined as the
time between formation and arrival in the ground state. The exact definition of 
the measured partial cascade time is contained in a non-trivial way in the data analysis
program of~\cite{pohl01}; we can approximate it by defining a collision to
be ``significant''  if the relative change in \mup\ kinetic energy is larger than
$\delta$ which we vary in the range $0.1-0.3$. 
The calculated cascade time and partial cascade time are 
shown in Figure~\ref{fig:mup:time} in comparison  
with the experimental data \cite{pohl01}. 
Whereas the cascade time grows as the pressure
decreases and eventually reaches the purely radiative cascade time, the partial
cascade time reaches its maximum at $2\cdot 10^{-7}\;$LHD and 
 vanishes at zero pressure where there are no collisions. 
The calculated partial cascade time  depends only weakly  
on $\delta$, and our results are in  good agreement with the experiment.

\begin{figure}
\center{
\epsfig{file=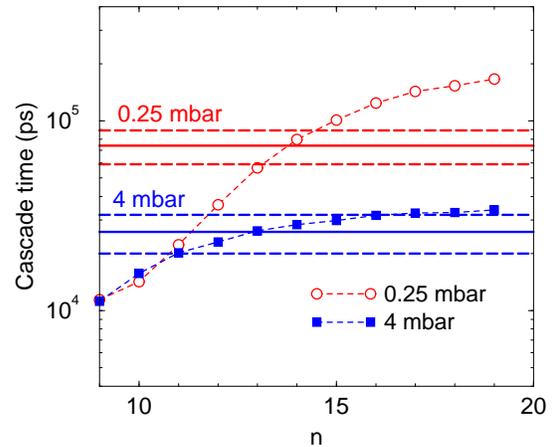, width=7.1cm}
}
\caption{\label{fig:mup:timen} 
The partial cascade time  as a function of the initial $n$ at
0.25~mbar and 4~mbar. The average initial energy is 0.5~eV and the parameter $\delta=0.1$.
The experimental data are  from \protect\cite{pohl01}.  
}
\end{figure}

The $n$ dependence of the partial cascade time at 0.25 and 4 mbar is shown in
Figure~\ref{fig:mup:timen}. At 4 mbar, the partial cascade time depends only weakly on the
initial $n$ and is in perfect agreement with the experimental result for $n>10$. The
calculated partial cascade time at 0.25~mbar has a moderate dependence on $n$ and agrees
with the experiment for $n=13-15$. 
Though this seems to support the simple picture of the \mup\ atoms being formed with
$n\sim 14$, the models that predict lower values for the initial $n$ can also be consistent
with the experiment because the $n$ distributions usually have tails that extends up
to high values. For example, using the  the $n$ distribution for molecular target
from~\cite{cohen99} we obtain the partial cascade time $\tau_\mathrm{casc}=60\;$ns 
at 0.25~mbar
in good agreement with the measured  $\tau_\mathrm{casc}=74\pm 15\;$ns\cite{pohl01}.

 
\subsubsection{Kinetic energy distribution}
\label{sect:res:mup:kin}

\begin{figure}
\center{
\epsfig{file=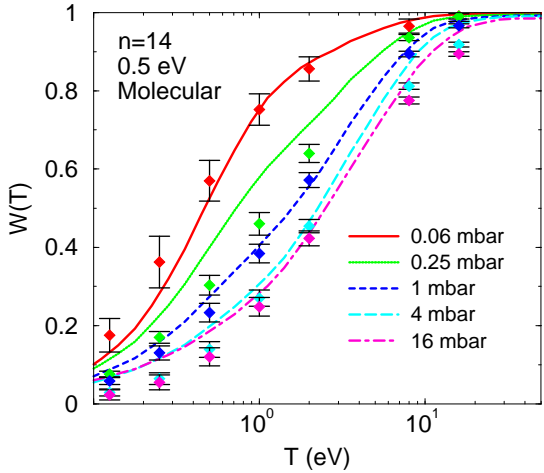, width=7.1cm}
}
\caption{\label{fig:mup:we14} 
The integrated energy distribution $W(T)$ of the \mup\ atom at the end 
of the cascade for  initial conditions: $n_i=14$ and $T_0 =0.5\;$eV.  
The data are from \protect\cite{pohl01}.  
}

\end{figure}

\begin{figure}
\center{
\epsfig{file=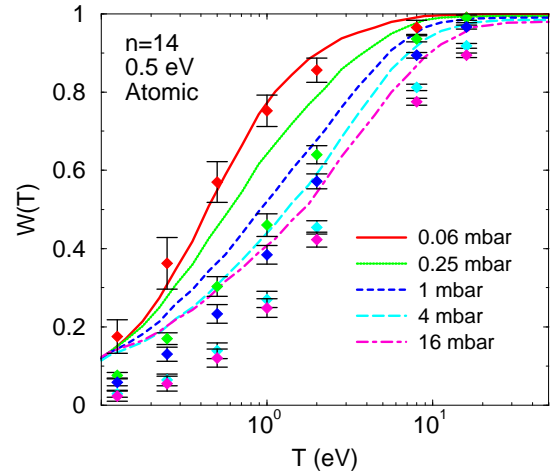, width=7.1cm}
}
\caption{\label{fig:mup:we14atomic} 
The integrated energy distribution $W(T)$ of the \mup\ atom at the end 
of the cascade for  initial conditions: $n_i=14$ and $T_0 =0.5\;$eV.  
The classical-trajectory calculations at $n>7$ were done for
{\it  atomic}  target. 
The data are from \protect\cite{pohl01}.  
}

\end{figure}

The kinetic energy distribution of muonic hydrogen  atoms  on arrival
in the $1s$ state has been obtained by analyzing \mup\ diffusion times in 
hydrogen gas~\cite{abbot97,kottmann99,pohl01,kottmann02hyp}.  
The integrated kinetic energy distribution\footnote{The reason for using the integrated instead
of the differential kinetic energy distribution, $w(T)$, is that neighboring energy bins
are highly correlated in the experimental analysis.} of muonic hydrogen on arrival
in the $1s$ state 
\be
W(T)=\int_{0}^{T}w(T^\prime)\dd T^\prime
\ee
was determined recently at PSI~\cite{pohl01,kottmann02hyp}.
  The results obtained in the pressure range
$0.06 -16$~mbar offer a unique possibility to study the formation of the \mup\ and
the initial stages of the cascade. Figure~\ref{fig:mup:we14} shows the integrated
kinetic energy distribution calculated in the cascade model and the experimental data 
from~\cite{pohl01}. 
  The data at 0.06~mbar are sensitive mainly to the initial energy distribution.  
With the density increasing, the role of the collisional processes grows,
and  the final distribution becomes more energetic 
due to acceleration at the beginning of the cascade. The results of the cascade 
calculations are in a fair agreement with the data for the initial conditions
specified in Section~\ref{sect:init}. 
The Coulomb transitions with $\Delta n>1$ were found to be essential  
for explaining the observed  increase in kinetic energy with increasing pressure
in agreement with the earlier analysis \cite{kottmann99}. Similar cascade calculations
using the trajectories of \mup\ collisions with  {\it atomic} 
hydrogen, where the collisions with $\Delta n=1$ dominate the collisional
deexcitation, predict a much weaker increase in kinetic energy as shown 
in  Figure~\ref{fig:mup:we14atomic}.



The cascade calculations predict, independent of the initial $n$,
significantly smaller fractions of $(\mup)_{1s}$ atoms with
high kinetic energies ($T\ge 8$~eV) at pressures above 4~mbar  
than the experimental data.



\begin{figure}
\center{
\epsfig{file=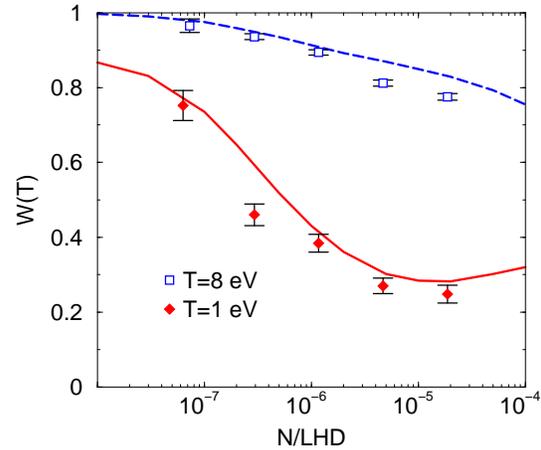, width=7.1cm}
}
\caption{\label{fig:mup:w1_8eV} 
The density dependence of  the integrated energy distributions 
$W(1\;\eV)$ and $W(8\;\eV)$ of the \mup\ atom
 for  initial conditions: $n_i=14$ and $T_0 =0.5\;$eV.  
The data are from \protect\cite{pohl01}.  
}

\end{figure}

Figure \ref{fig:mup:w1_8eV} shows the density dependence of 
$W(1\;\eV)$ and $W(8\;\eV)$.
 The calculated sub-1~eV  fraction, $W(1\;\eV)$, decreases from 80\% at $0.06\;$mbar 
to 30\% at 16 mbar in good agreement with the experimental 
data~\cite{pohl01}. The \mup\ atoms with energies above $8\;\eV$ are
produced during the cascade in Coulomb transitions. The present calculations predict
an increase in $W(T\ge 8\;\eV)$ with the pressure from 1\% at $0.06\;$mbar to 
17\% at $16\;$mbar. The predicted curves agree with the experimental results in the range
$0.06-1\;$mbar whereas the measured high energy fraction is substantially larger than
calculated one for pressures above $4\;$mbar: for example
$W(T\ge 8\;\eV)=(22.5\pm 0.9)\%$ at 16~mbar~\cite{pohl01}.
The measured increase in the high energy fraction of 
$(\mup)_{1s}$ atoms cannot be reproduced in the current cascade model
for any initial distribution.


\begin{figure}
\center{
\epsfig{file=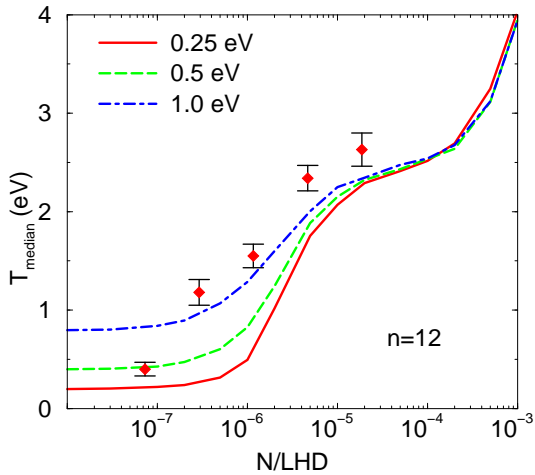, width=7.1cm}
}
\caption{\label{fig:mup:tmedian12} 
 The density dependence of the median kinetic energy of the \mup\ at the 
end of the cascade for different initial average kinetic energies and
$n_i=12$. 
The data are from \protect\cite{pohl01,pohl01median}.  
}
\end{figure}
\begin{figure}
\center{
\epsfig{file=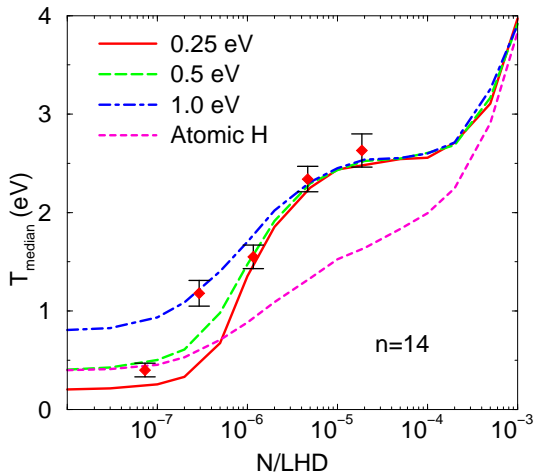, width=7.1cm}
}
\caption{\label{fig:mup:tmedian14} 
 The density dependence of the median kinetic energy of the \mup\ at the 
end of the cascade for different initial average kinetic energies and
$n_i=14$. The result for atomic target is shown for $T_0=0.5\;$eV.
The data are from \protect\cite{pohl01,pohl01median}.  
}
\end{figure}
\begin{figure}
\center{
\epsfig{file=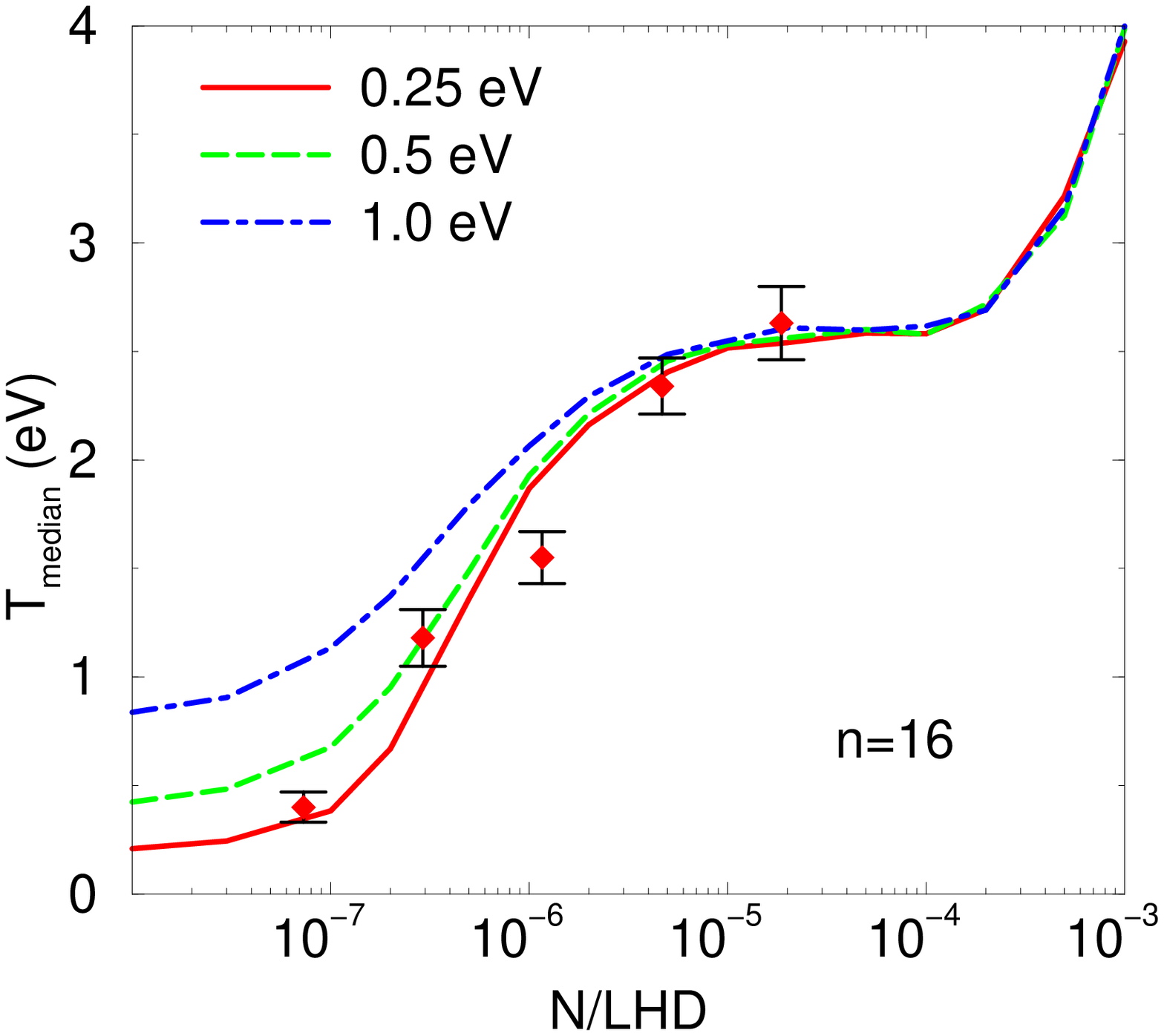, width=7.1cm}
}
\caption{\label{fig:mup:tmedian16} 
 The density dependence of the median kinetic energy of the \mup\ at the 
end of the cascade for different initial average kinetic energies and
$n_i=16$. 
The data are from \protect\cite{pohl01,pohl01median}.  
}
\end{figure}

Figures~\ref{fig:mup:tmedian12}, \ref{fig:mup:tmedian14}, and \ref{fig:mup:tmedian16} 
show the density dependence
of the median kinetic energy on arrival in the ground state for different
initial conditions. With the density increasing the calculated
median energies grow and reach $2.5\;\eV$ around $10^{-5}\;$LHD. 
Above $10^{-5}\;$LHD, the median 
energies remain  nearly constant up to $2\cdot 10^{-4}\;$LHD where they start to
grow again. The shape of the curves can be explained as follows. The increase
in the median energies
at low densities is produced in the classical domain ($n>7$)
where the acceleration due to Coulomb transitions
with $\Delta n=1-5$ is more efficient than the slowing down. The plateau is
due to the dominance of Auger deexcitation at medium $n$ which become 
important around the density $10^{-5}\;$LHD. The increase in the  median energy
in the uppermost part of the shown density range is produced by the Coulomb transitions
at $n<8$. 
For densities larger than $10^{-4}\;$LHD,
the initial kinetic energy is almost forgotten because of the many collisions
during the cascade,  and the kinetic energy distribution
at the end of the cascade is determined by the competition between acceleration
and deceleration during the cascade.

The measured density dependence of the median kinetic energy is well 
reproduced with the initial conditions:
$T_0=0.5\;$eV and $n_i=14$ or $T_0=0.25\;$eV and $n_i=16$. 
For  $n_i=12$ (Figure~\ref{fig:mup:tmedian12})  the initial acceleration appears to
be weaker and the agreement with the experimental result is worse.  
The molecular structure of the target is essential for explaining 
the data. Calculations with the  CMC trajectories  for  {\it atomic} hydrogen leads
to a too weak increase in the median energy compared to the experiment.

We estimate the statistical errors due to the finite number of CMC trajectories 
by dividing the 21 complete sets of trajectories into 7 subsets and calculate
the observables using each subset. The results for the median kinetic energy and the 
integrated kinetic energy shown in Table \ref{table:mup} does not change 
the conclusions reached above: the  ESCM gives a good description  of the experimental data for
moderate energies but is unable to explain the observed high energy fraction.
\begin{table}
\begin{center}
\begin{tabular}{|l|c|c|}\hline
 Observable &  ESCM  &  Exp.   
\\ \hline\hline
 $T_\mathrm{median}(1\;\mathrm{mbar})$& $1.57\pm0.03\;\eV$ &$1.55\pm0.12\;\eV$
\\ \hline
 $T_\mathrm{median}(16\;\mathrm{mbar})$& $2.50\pm0.14\;\eV$ &$2.63\pm0.17\;\eV$
\\ \hline
 $W(16\;\eV)$ at $16\;\mathrm{mbar}$ & $0.958\pm0.004$ &$0.894\pm0.006$
\\ \hline
\end{tabular}
\end{center}
\caption{\label{table:mup}
Median kinetic energy and integrated kinetic energy in muonic hydrogen 
calculated in the ESCM 
with statistical errors. The experimental results are from \protect\cite{pohl01,pohl01median}.}
\end{table}

\subsection{Antiprotonic hydrogen}
   The present study of the atomic cascade in antiprotonic hydrogen was 
focused on the experimental data obtained at low density 
\cite{obelix00,bacher89,baker88,heitlinger92,eijk88,gotta99}.

\subsubsection{X-ray yields}

\begin{figure}
\center{
\epsfig{file=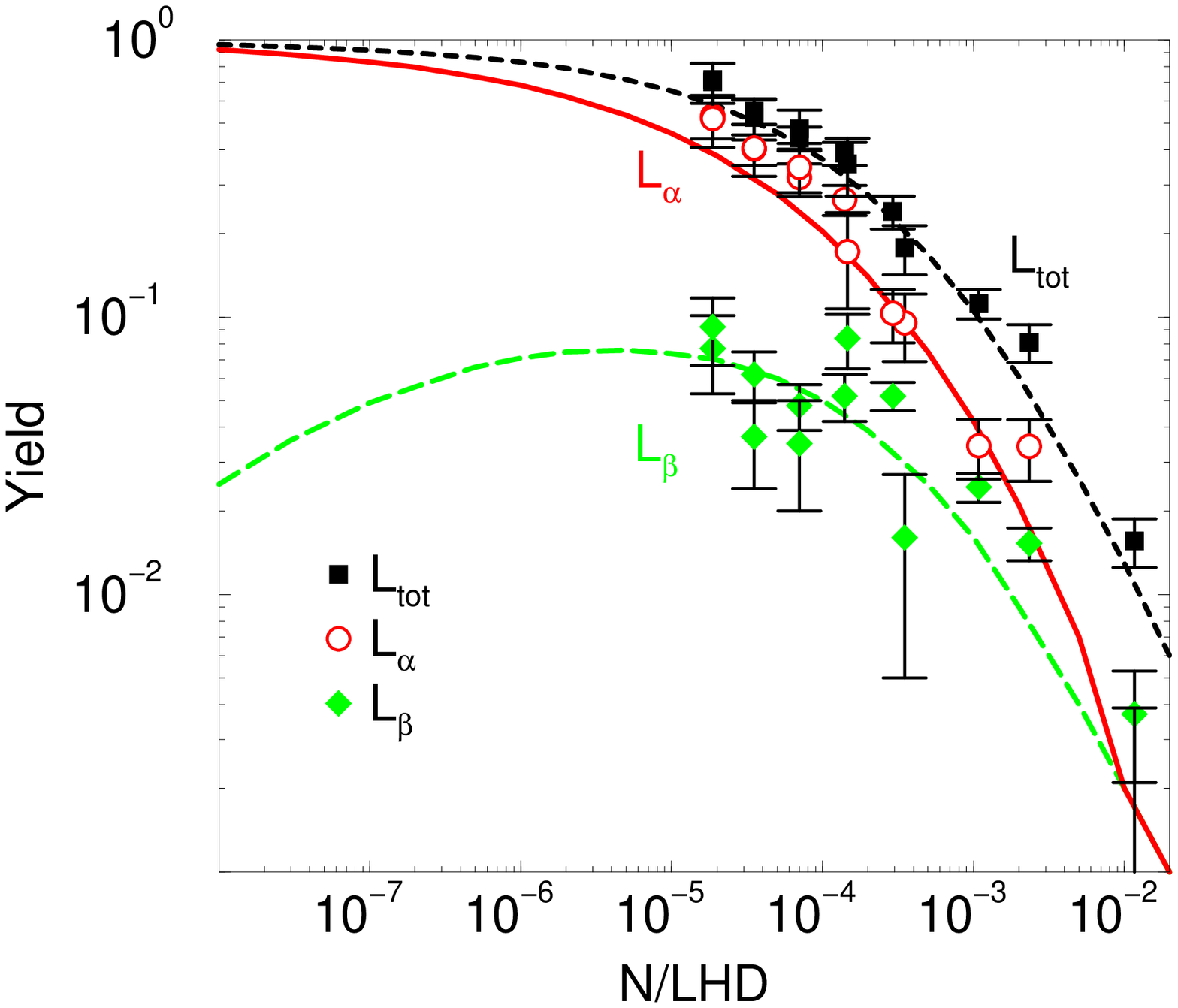, width=7.1cm}
}
\caption{\label{fig:pbp:Lxray}
 The density dependence of the absolute $L$  X-ray yields 
in antiprotonic hydrogen. 
The experimental data are from
\protect\cite{bacher89,baker88,heitlinger92,eijk88}.  
}
\end{figure}
\begin{figure}
\center{
\epsfig{file=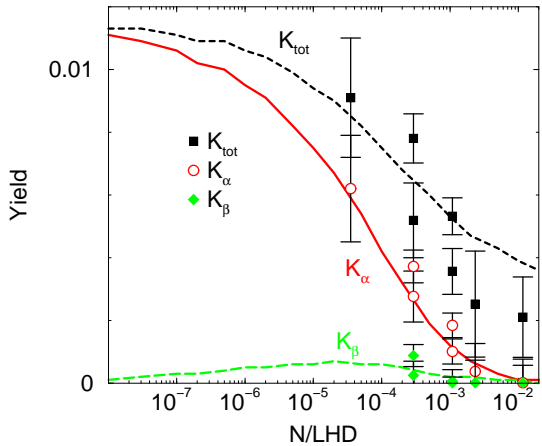, width=7.1cm}
}
\caption{\label{fig:pbp:Kxray}
 The density dependence of the absolute 
 $K$ X-ray yields 
in antiprotonic hydrogen. 
The experimental data are from
\protect\cite{bacher89,baker88,heitlinger92,eijk88}.  
}
\end{figure}

\begin{figure}
\center{
\epsfig{file=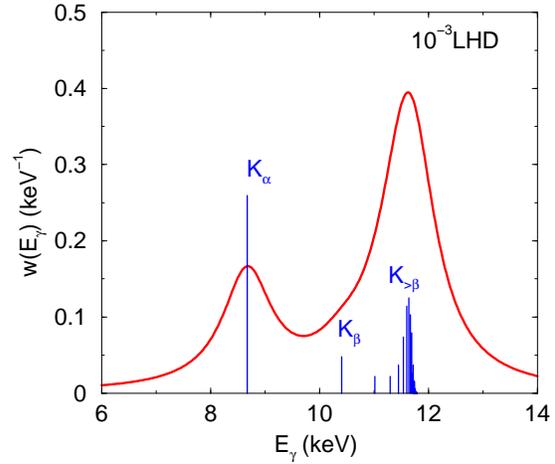, width=7.1cm}
}
\caption{\label{fig:pbpprofile}
The $K$ line profile in antiprotonic hydrogen at $10^{-3}\;$LHD.}
\end{figure}

  The calculated X-ray yields are in 
a good agreement with the data \cite{bacher89,baker88,heitlinger92,eijk88} 
as shown in Figures~\ref{fig:pbp:Lxray} and~\ref{fig:pbp:Kxray}. 
In addition to the data on absolute X-ray yields, there are more precise measurements
of relative yields, see Table~\ref{table:pbpyields}. The theoretical results
are in fair agreement with the data as all the calculated ratios deviate less
than 50\% from the experimental results. This is a significant  improvement in comparison
with  the calculations in the Borie-Leon model reported in~\cite{heitlinger92}: 
the two standard tuning  parameters \kstk\ and $T$ were fixed by fitting the absolute
$L_\alpha$ yields but the predictions of the model for ratios involving 
$L_\delta$ were a factor of $3-5$ larger than the experimental data.
We stress that our results were obtained without employing any  cascade 
tuning parameters.      
Figure~\ref{fig:pbpprofile} shows an example of the X-ray  profile of the $K$ lines at
$10^{-3}\;$LHD. 
Because of the hadronic broadening of the $1s$ state, 
only the $K_\alpha$ line is clearly separable from the rest of the $K$ lines of which
the $9p\to 1s $ transition has the highest intensity. The  
relatively large contribution to the $K$ yield from high $n$ states is due to the 
strong absorption in the excited states which  prevents most of the \pbp\ atoms from reaching 
the low $n$ states.

\begin{table*}
\begin{center}
\begin{tabular}{|l|c|c|c|c|}\hline
 Observable & Pressure  & ESCM  &  Exp.  &  Reference  
\\ \hline\hline
$Y(K_\alpha)/Y(L_\alpha)$ & 20 mbar  & 0.0179& $0.0176\pm0.0016 $ & \cite{augsburger99} 
\\ \hline
$Y(L_\mathrm{tot})/Y(L_\alpha)$ & 20 mbar& 1.56 & $1.45\pm 0.05 $  & \cite{augsburger99} 
\\ \hline
$Y(L_\alpha)/Y(L_\beta)$ & 30 mbar& 5.0 & $7.1\pm 0.7$  & \cite{heitlinger92} 
\\ \hline
$Y(L_\alpha)/Y(L_\gamma)$ & 30 mbar& 13.2 & $20.6\pm 2.0$  & \cite{heitlinger92} 
\\ \hline
$Y(L_\alpha)/Y(L_\delta)$ & 30 mbar&  22.6 & $23.9\pm 2.6$  & \cite{heitlinger92} 
\\ \hline
$Y(L_\alpha)/Y(L_\mathrm{tot})$ & 30 mbar& 0.62 & $0.75\pm 0.08$  & \cite{heitlinger92} 
\\ \hline
$Y(L_\beta)/Y(L_\gamma)$ & 30 mbar&  2.7 & $2.9\pm 0.2$  & \cite{heitlinger92} 
\\ \hline
$Y(L_\beta)/Y(L_\delta)$ & 30 mbar& 4.6 & $3.4\pm 0.2$  & \cite{heitlinger92} 
\\ \hline
$Y(L_\gamma)/Y(L_\delta)$ & 30 mbar& 1.7 & $1.2\pm 0.1$  & \cite{heitlinger92} 
\\ \hline
\end{tabular}
\end{center}
\caption{\label{table:pbpyields}
Relative X-ray yields in \pbp.}
\end{table*}

\subsubsection{Cascade time}

\begin{figure}
\center{
\epsfig{file=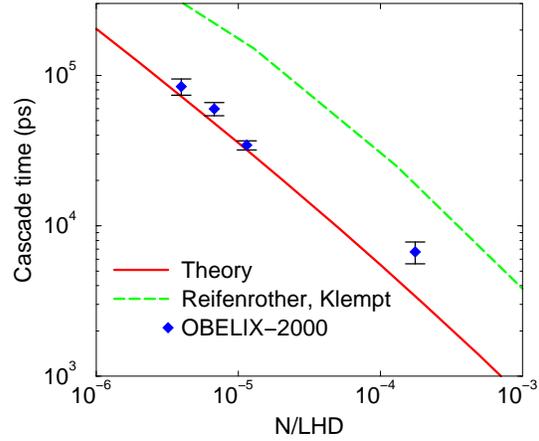, width=7.1cm}
}
\caption{\label{fig:pbp:time}
 The density dependence of the cascade time in antiprotonic hydrogen.
    The experimental data are from \protect\cite{obelix00}.  
}
\end{figure}
\begin{figure}
\center{
\epsfig{file=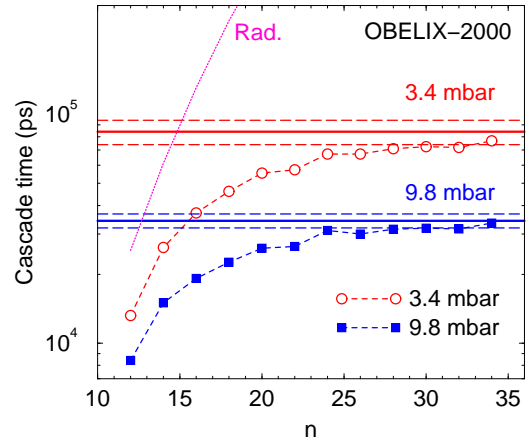, width=6.8cm}
}
\caption{\label{fig:pbp:timen}
 The dependence of the cascade time in \pbp\ at 3.4 and 9.8~mbar on the initial 
    state $n$. The initial average kinetic energy is 0.5~eV. The cascade time for 
    a purely radiative cascade is shown for comparison.
    The experimental data are from \protect\cite{obelix00}.  
}
\end{figure}

   The density dependence of the cascade time is shown in 
Figure~\ref{fig:pbp:time}.  
Our result is in good agreement with the data \cite{obelix00} except for 
the data point at 150~mbar where our model predicts significantly shorter
cascade time. The agreement for pressures up 
to 10~mbar  can 
be considered as an important confirmation of the dominance of the Coulomb 
transitions with large changes in  the principal quantum number $n$ in  the 
upper part of the atomic cascade.  
The earlier calculations \cite{reifenrother89} predicted much higher 
cascade times in variance with the experimental data.

   
The dependence of the cascade time at 3.4 and 9.8~mbar on the initial $n$ is shown 
in Figure~\ref{fig:pbp:timen}. There is good agreement 
with the experimental result \cite{obelix00} in the 
range $n=25-35$. Calculations also show that the 
calculated cascade times depend  only weakly on the initial kinetic energy.

\subsubsection{Kinetic energy distributions}

\begin{figure}
\center{
\epsfig{file=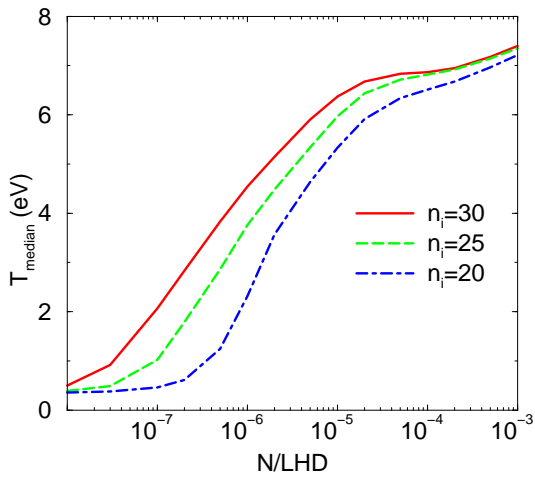, width=7.1cm}
}
\caption{\label{fig:pbp:tmedian}
 The density dependence of the median energy at the instant of nuclear
absorption in antiprotonic hydrogen for initial $n=20,25,30$. 
}
\end{figure}

In the upper part of the cascade most of the \pbp\ atoms are  accelerated to 
energies of several eV in the Coulomb transitions with $\Delta n>1$. 
Figure~\ref{fig:pbp:tmedian}
shows the density dependence of the median energy at the instant of nuclear
absorption calculated for initial $n=20,25,30$. 
The shape of the curves are similar to those of muonic hydrogen but the energy is higher.
The median energy increases rapidly with increasing density and
reaches  $6-7\;\eV$ at
$10^{-5}\;$LHD. At higher densities the  dominance of the Auger transitions  at intermediate
$n$ prevents further increase until  the Coulomb deexcitation at low $n$ becomes important. 


\subsubsection{Doppler broadening of the $L$ lines}

\begin{figure}
\center{
\epsfig{file=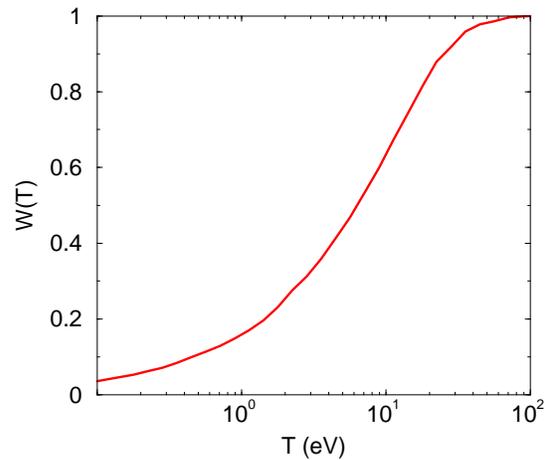, width=7.1cm}
} 
\caption{\label{fig:pbp:we22mbar} 
 The calculated integrated energy distribution of \pbp\  
    at the instant of the $3d\to 2p$ transition at 22~mbar.  
}
\end{figure}
\begin{figure}
\center{
\epsfig{file=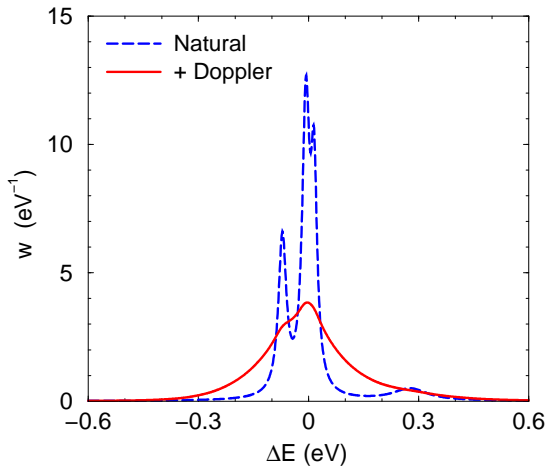, width=7.1cm}
} 
\caption{\label{fig:pbp:doppler} 
 The natural $L_{\alpha}$ line profile  in \pbp\
    (dashed line) in comparison with the  Doppler broadened line profile 
    at 22~mbar (solid line). For the electromagnetic hyperfine splitting 
    we use the  result of Boucard and Indelicato~\protect\cite{gotta99}. 
    The hadronic splitting and  broadening are taken from the DR1 model
    of~\protect\cite{carbonell89}.
}
\end{figure}

  The acceleration processes during the atomic cascade 
lead to a sizeable Doppler broadening of the $L$ lines. 
Figure~\ref{fig:pbp:we22mbar} shows the integrated energy 
distribution of the \pbp\ atoms at the instant of the $3d\to 2p$ transition 
in a gaseous target of 22~mbar, the median energy being about 6.5~eV.  
The corresponding Doppler broadening of the $3d\to2p$ line profile 
is  shown in Figure~\ref{fig:pbp:doppler}; 
the effect gives a significant contribution  
to the width and must, therefore, be taken into account in the
analysis of the X-ray spectra. In the data analysis in~\cite{gotta99}
it was assumed that the \pbp\ atoms were thermalized due to elastic and
Auger collisions.

\section{Conclusions}
\label{sect:conc}


Using the new results for the collisional processes~\cite{jensen02epjd,jensen02last},
the atomic cascades in muonic and antiprotonic hydrogen have been  
have been studied in detail. The cascade calculations have been done in the extended standard
cascade model which describes the evolution in the quantum numbers $n$ and $l$ and the
kinetic energy from the very beginning of the cascade. By taking deceleration and 
acceleration during the cascade into account, observables can be predicted reliably
without the need for any cascade tuning parameters. The extended standard
cascade model successfully describes a number of experimental results: X-ray yields,
cascade times, and kinetic energy distributions. 
The results of the cascade calculations for muonic and antiprotonic
hydrogen  can be summarized as follows:
\begin{itemize}
\item{Muonic hydrogen. The calculated X-ray yields in muonic hydrogen are in good agreement with the data. 
       The measured ratio of the $K_\alpha$ and  $K_\beta$ yields in liquid hydrogen 
       can be reproduced only if the non-radiative quenching of the metastable 
       $2s$ state is dominant.
        The calculated  increase in the $(\mup)_{1s}$ median kinetic energy 
       with the density is in agreement with the data~\cite{pohl01} for the initial 
       conditions corresponding to 
       kinetic energies around 0.5~eV and principal  quantum number $n\approx 14$. 
       The molecular structure of the target is essential for explaining the data
       on the density dependence of the kinetic energy distribution at the end of 
       the cascade.   
 }
\item{Antiprotonic hydrogen. The calculated $K$ and $L$ X-ray yields   in antiprotonic hydrogen are in good agreement 
       with the data.  The calculated cascade time is in good agreement with the data. 
       For the first time, a good description of the data is obtained without 
       using any phenomenological tuning parameters. 
       The fast collisional deexcitation at high $n$ calculated in the  
       classical-trajectory  model was found to be  very important for explaining the 
       short cascade times measured by the OBELIX collaboration~\cite{obelix00}. The acceleration
       of the \pbp\ during the initial stages of the cascade predicted by our cascade model 
       results in  a significant Doppler broadening of the $L$ lines. This must be taken
       into account when hadronic widths and shifts are obtained from the measured line profiles. 
}
\end{itemize}
 
Some experimental data cannot be explained in the present version
of the ESCM even with taking the uncertainties in the cross-sections into account. 
The  problem of the high energy fraction of the kinetic energy distribution
in muonic hydrogen deserves special attention in future studies: 
the ESCM predicts a much smaller fraction  
of \mup\ atoms with energies above 16 eV at 4 and 16 mbar 
than the experimental result~\cite{pohl01}. 
This disagreement indicates that some processes beyond the ESCM are important.  
Candidates under discussion are 
the hybrid Auger-Coulomb transitions~\cite{faifman02hyp} and
the resonant formation of molecular states~\cite{jonsell99}.

Future development of the ESCM will consist of calculations of the 
above mentioned processes and their inclusion in the cascade model. 
At the same time the cascade calculations can be easily extended to cover 
exotic deuterium atoms and H-D mixtures. 
Further critical tests of the current theory of the atomic cascade 
will greatly benefit from the results of the experiments on the 
precision spectroscopy of pionic and muonic hydrogen, presently in 
progress at PSI.  


\section*{Acknowledgment}

We thank F.~Kottmann, L.~Simons, D.~Taqqu, R.~Pohl, and 
D.~Gotta for fruitful and stimulating discussions.


\bibliographystyle{unsrt}

\end{document}